\documentstyle[prb,eqsecnum,floats,aps,twocolumn]{revtex}
 
\input{psfig}
\begin{document}
\draft
\twocolumn[\hsize\textwidth\columnwidth\hsize\csname @twocolumnfalse\endcsname 
 
\title{ Contacts and Edge State Equilibration in the Fractional
Quantum Hall Effect}
 
\author{C.L. Kane}
 
\address{Department of Physics, University of Pennsylvania\\
Philadelphia, Pennsylvania 19104
}
\author{Matthew P.A. Fisher}
\address{Institute for Theoretical Physics, University of California\\
Santa Barbara, CA 93106-4030
}
 
\date{\today}
\maketitle
 
\begin{abstract}
We develop a simple kinetic equation description
of edge state dynamics in the fractional quantum Hall
effect (FQHE), which allows us to examine in detail
equilibration processes between multiple edge modes.
As in the integer quantum Hall effect (IQHE), inter-mode equilibration is a prerequisite for quantization of the Hall
conductance.  Two sources for such equilibration
are considered:
Edge impurity scattering and equilibration by the electrical contacts.
Several specific models for electrical contacts
are introduced and analyzed.  For FQHE states in which edge channels move
in both directions, such as $\nu=2/3$, these models for the electrical contacts
{\it do not} equilibrate
the edge modes, resulting in a non-quantized Hall
conductance, even in a four-terminal measurement.
Inclusion of edge-impurity scattering, which {\it directly} transfers
charge between channels,
is shown to restore the four-terminal quantized conductance.
For specific filling factors, notably $\nu =4/5$ and $\nu=4/3$,
the equilibration length due to impurity scattering diverges
in the zero temperature limit, which should lead to a breakdown
of quantization for small samples at low temperatures.
Experimental implications are discussed.

\end{abstract}


]

\section{Introduction}

An important lesson learned from studies
of mesoscopic structures is 
that the transport properties of a system can be
strongly influenced by the electrical 
contacts used to make the measurements\cite{review}.
The nature of the contacts can be
particularly important in the
quantum Hall regime, where transport takes place
via edge states.  Current carrying contacts feeding the edge states
can alter their population, leaving the edge out of equilibrium.
High impedance voltage contacts selectively measure the electrochemical
potential of the edge modes to which they are most strongly coupled.

In the integer quantum Hall effect, B\"uttiker\cite{Buttiker} has generalized
Landauer quantum transport theory\cite{Landauer} to
incorporate
the effects of multiple electrical contacts.
A contact, or lead, is modeled as a large reservoir in equilibrium at an
electro-chemical potential $\mu$.  
The edge states in the sample are populated by electrons incident
from the leads.
B\"uttiker defines an ``ideal" contact as one in which
no scattering occurs at the contact.   The
edge states emanating from such ideal contacts are thus
populated in equilibrium at the chemical potential $\mu$.
B\"uttiker also discusses the more generic case of a disordered or
``non ideal" contact, which  is characterized by
transmission and reflection matrices between
the edge channels in the ``sample" and in the ``leads".  

At integer filling factors $\nu > 1$, there are multiple
edge channels.  For generic ``non-ideal" contacts, the coupling
to the different edge modes will be different.
Non-ideal current contacts will thus tend to populate
the edge modes differently, putting them out of equilibrium with one
another.
In this case, the edge is {\it not} characterized
by a unique chemical potential, and the Hall conductance measured using 
similar non-ideal voltage contacts will {\it not} be quantized.  
For this reason, 
B\"uttiker emphasizes the important role played by inter-channel
electron scattering, 
which can
re-equilibrate the different modes.  Provided the separation between
current and voltage leads is greater than the equilibration
length, the voltage lead will measure an equilibrated edge,
giving a 
quantized
Hall conductance.

By fabricating non-ideal contacts which are close together,
it is possible to study directly the equilibration between edge states.
Indeed, in beautiful experiments utilizing a quantum point contact
which selectively populates channels\cite{Alphenaar,Kouwenhoven,Takagaki}, 
equilibration lengths of order $40 \mu m$
have been measured in the integer quantum Hall regime.

While the importance of contacts and
edge state equilibration is well appreciated for the integer
quantum Hall effect, a suitable generalization to the
fractional quantum Hall regime has been lacking.  
In the fractional quantum Hall effect (FQHE), the edge modes
cannot be described in terms of a free electron model,
so that a Landauer-B\"uttiker description
of edge transport is not possible.  Recently a
powerful framework for describing  
edge states in the FQHE has been developed, based on the
chiral Luttinger liquid model\cite{Wen}.
This description enables one to compute
transport properties
of edge states in the FQHE.  Recently, the effects of impurity scattering
on edge state equilibration has been discussed in the FQHE\cite{KFP,sun}, 
but the role of electrical contacts
has not been adequately addressed\cite{Haldane2}. 

In this paper we develop a simple theory for edge state transport
in the FQHE which allows for incorporation of electrical contacts
and inter-mode equilibration.
The approach is
based on a simple kinetic equation for the edge state dynamics, 
which closely resembles a linearized
Boltzmann equation.  Coupling to electrical contacts
is incorporated by adding source terms to the kinetic equation, loosely analogous
to scattering terms in the Boltzmann equation.
Impurity scattering between multiple edge modes can also be simply
incorporated into the kinetic equation.
Several specific models for the contacts are considered,
which are analogous to B\"uttiker's ``ideal"
and ``non ideal" contacts in the IQHE.    

For simplicity we focus on quantum Hall states at 
filling $\nu = (p_1 - 1/p_2)^{-1}$  
($p_1$ odd, $p_2$ even), which correspond to the second level
of the Haldane-Halperin hierarchy\cite{HaldaneHalperin}
 and have two edge channels.
As in the integer quantum Hall effect, we find that 
equilibration between the different
edge channels is a prerequisite for the quantization of the
Hall conductance.
There are two sources for this equilibration which we address
separately:  Impurity scattering along the edge and equilibration
by the contacts themselves.  In the former case, we find important differences
with the IQHE.  Specifically, for 
filling fractions with $|p_2| = 4,6,...$, such as $\nu = 4/3,4/5,...$,
the inter-mode equilibration length due to impurity scattering
is temperature dependent
and {\it diverges} at low temperatures.  In this case, for
finite sized sample, quantization
of the Hall conductance should break down at very low
temperatures. 

There are also important differences between
the integer and fractional Hall effects with regards the equilibration taking 
place {\it at} the contacts.
The differences are most pronounced when the
two fractional edge modes are moving in opposite directions, for
$p_2<0$ at filling $\nu = 2/3, 4/5,...$.
In this case, even ''ideal contacts" are insufficient
to equilibrate the two edge modes.  The reason is that 
the two modes emanate from different reservoirs,
at different chemical potentials.  Then in the absence of any direct
inter-mode impurity tunneling, the two channels on the same edge
will be at different chemical potentials.
Impurity scattering away from the contacts can still cause
equilibration, but with an equilibration length diverging
at low temperatures for $|p_2|\ge 4$.
In contrast, when both edge modes propagate in the same
direction ($p_2 >0$), we show that ``ideal contacts" can be defined
which completely equilibrate the two channels, just as in the IQHE.

In this paper we focus almost exclusively on the linear
response conductances, for a sample with {\it finite}
width, $L$.  Within linear response, both the Hall voltage
drop $V_H$, and the Hall electric field $E=V_H/L$,
are taken to zero, with {\it fixed} width $L$.  
In this limit, the edge currents give an order
one contribution to the Hall conductance,
and with long-ranged forces screened
by a ground plane, dominate completely over bulk contributions.
As we shall see, the conductance in this case 
is a ``mesoscopic" quantity, which can
depend on the nature of the electrical contacts and 
of the edge states which transport
current between them. 
In contrast, it is possible to define a ``macroscopic" Hall conductance
which is a bulk property, independent of the edge dynamics.
In the limit that $L \rightarrow \infty$,
with fixed finite electric field, $E$, the edge contribution to the
bulk Hall conductance vanishes as $1/L$.

Our paper is organized as follows.  In section II,
we introduce the simple kinetic equation description of FQHE
edge state transport, for the case in which only a single edge mode
is expected, $\nu^{-1}$ an odd integer. 
Various models for electrical contacts are discussed
within this framework.  In section III, the description is generalized
to describe hierarchical Hall states with two edge modes.
Equilibration between the two edge modes both at the contacts, and due
to edge impurity scattering, is discussed.
In section IV, we describe several specific experimental consequences,
and conclude in Section V.

\section{SINGLE EDGE MODE}

In this section we introduce a simple kinetic equation description
for the edge of a Laughlin state at filling $\nu = 1/m$.
In this case there is only a single edge mode, which satisfies
a simple continuity equation.  We then show how electrical contacts
can be incorporated into this approach, and consider specific
models for the contacts.  Finally, we show how bulk electric fields,
and bulk currents, can also be incorporated, without
changing the conclusions.  In Section III we will turn our attention to
hierarchical Hall states, which have multiple edge modes.

\subsection{Kinetic Equation}

For filling $\nu=1/m$ with odd integer $m$,
a single chiral edge mode is expected\cite{Wen}. 
 The 1d electron density, $n(x)$,
satisfies a simple equation of motion,
\begin{equation}
\partial_t n + v \partial_x n =0 ,
\end{equation}
which describes waves moving in one direction at velocity $v$:
$n(x,t) = f(x-vt)$, for arbitrary $f$.
This can be written as a continuity equation,
\begin{equation}
\partial_t n + \partial_x J = 0 ,
\end{equation}
with an edge current defined as
\begin{equation}
J = v n .
\end{equation}
Eqn. (2.2) is a conservation law for electric charge at the edge.
Together, (2.2) and (2.3) are a simple kinetic 
equation for edge charge transport.

Since the bulk is incompressible, charge cannot pass from the edge into
the bulk, at least in linear response.  
In the presence of a large non-linear driving field, though,
bulk currents can flow, and it is necessary to augment eqn. (2.2).
Moreover, charge can be added or removed from the edge mode
at contacts.  In these cases source terms must be added
to the right side of (2.2):
\begin{equation}
\partial_t n + \partial_x J = I_{bulk}(x) + I_{c}(x)  ,
\end{equation}
which describe charge being added or removed
from the edge, via either bulk currents, or from contacts.    

The kinetic equation (2.4) is 
analogous to the transport equation for 
quasiparticles in a Fermi liquid\cite{Nozieres}. 
In the present case, the Fermi surface is replaced by a single point. 
The left hand side describes collisionless transport, 
and follows directly from  microscopic equations of motion,
as we show below.  The term on the right hand side coming
from the contacts is
analogous to a ``collision term" in the Boltzmann equation.
An explicit form for this term can be obtained by using Fermi's Golden
rule, as we show below, which is the rough equivalent of the relaxation
time approximation.
This treatment requires that the time between 
successive tunneling events from the contacts exceeds
the
dephasing time $\tau_\phi$.   
We shall describe the role of the bulk currents in section 2C.

We now show briefly that the collisionless terms (2.2) and (2.3)
follow directly from a chiral Luttinger
liquid\cite{Wen} description of the edge mode.
In terms of a boson field, $\phi$, related to the electron density
via $n(x)=(1/2\pi) \partial_x \phi$, the chiral Luttinger Hamiltonian
is simply:
\begin{equation}
H = {v\over {4\pi\nu}}  \int dx 
(\partial_x \phi)^2   ,
\end{equation}
where the phase field satisfies a Kac-Moody commutation relation:
\begin{equation}
[\phi(x),\phi(x^\prime)] = i \pi \nu {\rm sgn}(x-x^\prime) .
\end{equation}
>From the Heisenberg equations of motion for the operator $\phi$,
it is straightforward to show that the density operator satisfies the kinetic
equation (2.1). 
 
The continuity equation (2.2) can be re-written in 
the suggestive form:
$\partial_x (J + (\partial_t \phi /2\pi))=0$,
allowing us to identify the current operator as $J = -\partial_t \phi/2\pi$.
It is useful to assume normal ordering, so that in equilibrium
both the density and currents vanish.
In the presence of a non-zero chemical
potential, $\mu$, however, currents will flow.
The current response can be deduced by adding to the Hamiltonian
a term of the form:
\begin{equation}
\delta H = - \mu \int dx  n  ,
\end{equation}
and then evaluating the current, $J = -e \partial_t \phi/2\pi$,
using the commutation relations (2.6).
One deduces a non-vanishing transport current of the form:
\begin{equation}
J = \nu {e\over h} \mu  .
\end{equation}
The conductance is seen to be appropriately quantized, 
$G = eJ/\mu = \nu e^2/h$.

\subsection{Contacts}

We consider now incorporating electrical contacts into
the above description of edge transport.
We begin by discussing B\"uttiker's ``ideal" contact\cite{Buttiker}
 model generalized to 
the fractional quantum Hall regime.  While this model is useful
for performing simple calculations, it is rather unrealistic - particularly
in the FQHE - 
since it assumes that the edge modes
retain their integrity deep within the reservoirs.
As an alternative, we consider a contact modeled as 
a tunnel junction to a metallic electrode.  In B\"uttiker's
terminology, this is an example of a ``non ideal" contact.
The point contact tunnel junction can be suitably generalized,
however, 
into a tunnel junction ``line contact".  The ``line contact"
is shown to be ``ideal", with 
vanishing contact resistance.

\subsubsection{The Ideal Contact}

In the Landauer-B\"uttiker approach to quantum transport,
electrical contacts are modeled as reservoirs at chemical potential
$\mu$.  Transport is viewed as a scattering process.
Electrons incident from the reservoirs
enter the sample, scatter about and then leave the sample back
into the reservoirs.  In this scheme an ``ideal contact"
is one in which there is no electron
backscattering during the process of entering or leaving the sample.
The 
population of an each edge mode is then determined by the
chemical potential of the reservoir from which it emanates.

For an interacting system, such as a
FQHE edge mode, the transport can not be described in terms
of the population of free electron states.  Nonetheless,
a Landauer-type formula for transport
can be derived within a linear response Kubo formulation
\cite{FisherLee,Stone}.
The ``reservoir" is modeled
as a semi-infinite strip of FQHE fluid connected to the ``sample".  
The edge channels extend to infinity in the ``reservoirs",
as shown in Fig. 1a.  Following Fisher and Lee,
the conductance may then be computed within linear response
theory by applying a time dependent potential,
$V(x)e^{i\omega t}$, where $V(x)$ 
is equal to $\mu_i$ in the i'th reservoir, and then taking
the $\omega\rightarrow 0$ limit.
For non interacting electrons this procedure 
is equivalent to the Landauer approach, but can be suitably
generalized to the FQHE.
It results in a non-equilibrium current flowing in the fractional
 edge channels, determined by the chemical potential 
of the reservoir from which they emanate,
$J=\nu (e/h) \mu_i$.  If the Hall voltage is measured 
on the top and bottom edges by similar ``ideal contacts",
the result is an appropriately quantized Hall conductance.
Moreover, the two terminal conductance, 
defined as the ratio of the current to the source-drain
voltage, is also quantized.

The assumption that the edge modes maintain their 
integrity deep within the reservoirs is highly unphysical.
This feature is particularly worrisome in the FQHE, where the
the edge modes are gases
of fractionally charged quasiparticles. One would 
expect an appreciable contact resistance as the electrons from
the metallic electrodes splinter upon
entering the sample, in contrast to the resistanceless
``ideal contacts" considered above. 
We now consider more realistic models for electrical contacts,
consisting of a metallic electrode coupled to the edge modes via
tunnel junctions.

\subsubsection{The tunnel junction point contact}

Consider a metallic electrode, described by a Fermi liquid at
chemical potential $\mu$.
The electrode is connected to the edge mode via a tunnel junction,
at position $x=0$.  The tunneling process
transfers an electron from metallic electrode
to the edge with Hamiltonian:
\begin{equation}
H_{tun} = -t_0  \psi (x=0) \int dx \delta(x) e^{i \phi(x)/\nu} + h.c. .
\end{equation}
Here $\psi$ is an electron destruction operator in the metallic electrode,
and $e^{i\nu\phi}$ is the edge electron creation operator.
This tunneling Hamiltonian leads to an additional term on the 
right side of the continuity
equation (2.2),
\begin{equation}
\hat{I}_c = \delta(x) i t_0 \psi(x) e^{i\phi(x)/\nu} - h.c. ,
\end{equation}
which describes tunneling of charge from electrode to edge.
Since this operator is non-linear,
it is desirable to replace it by it's
expectation value, so that (2.2) can still be used
as a classical kinetic equation.
This simplification requires that   
successive tunneling events are incoherent.
In the limit $t_0 \rightarrow 0$ this should be the case,
since the time between tunneling events then exceeds 
the electron dephasing times 
both in the electrode and on the edge.
In this limit, the average tunneling current is given 
by a contact conductance, $G_{c}$,
times the chemical potential drop between electrode and the edge
mode.  The chemical potential of the edge mode can
be obtained from the upstream current entering the contact region:
$\mu_{edge} = J(x=0^-) h/ (\nu e) $.
Thus the kinetic equation can written as a closed expression in terms of
the edge density and current:
\begin{equation}
\partial_t n + \partial_x J = I_c(x) ,
\end{equation}
with a contact current:
\begin{equation}
I_c(x) = \delta(x) G_c \left(\mu - {h\over {\nu e}} J(x)\right)  .
\end{equation}
When many contacts are present, at positions $x_i$,
and chemical potentials $\mu_i$, this generalizes to
\begin{equation}
I_c(x) = \sum_i \delta(x_i) G_c \left(\mu_i - {h\over {\nu e}} J(x_i)\right) .
\end{equation}
The tunneling conductance will in general be temperature dependent.  
Using Fermi's Golden rule perturbative in the tunneling matrix element
$t_0$, gives
$G_{c} \sim t_0^2 T^{(1/\nu -1)}$, which vanishes at low
temperature in the FQHE.  We will assume in any event, that
$G_c << e/h$. 
\begin{figure}
\centerline{\psfig{figure=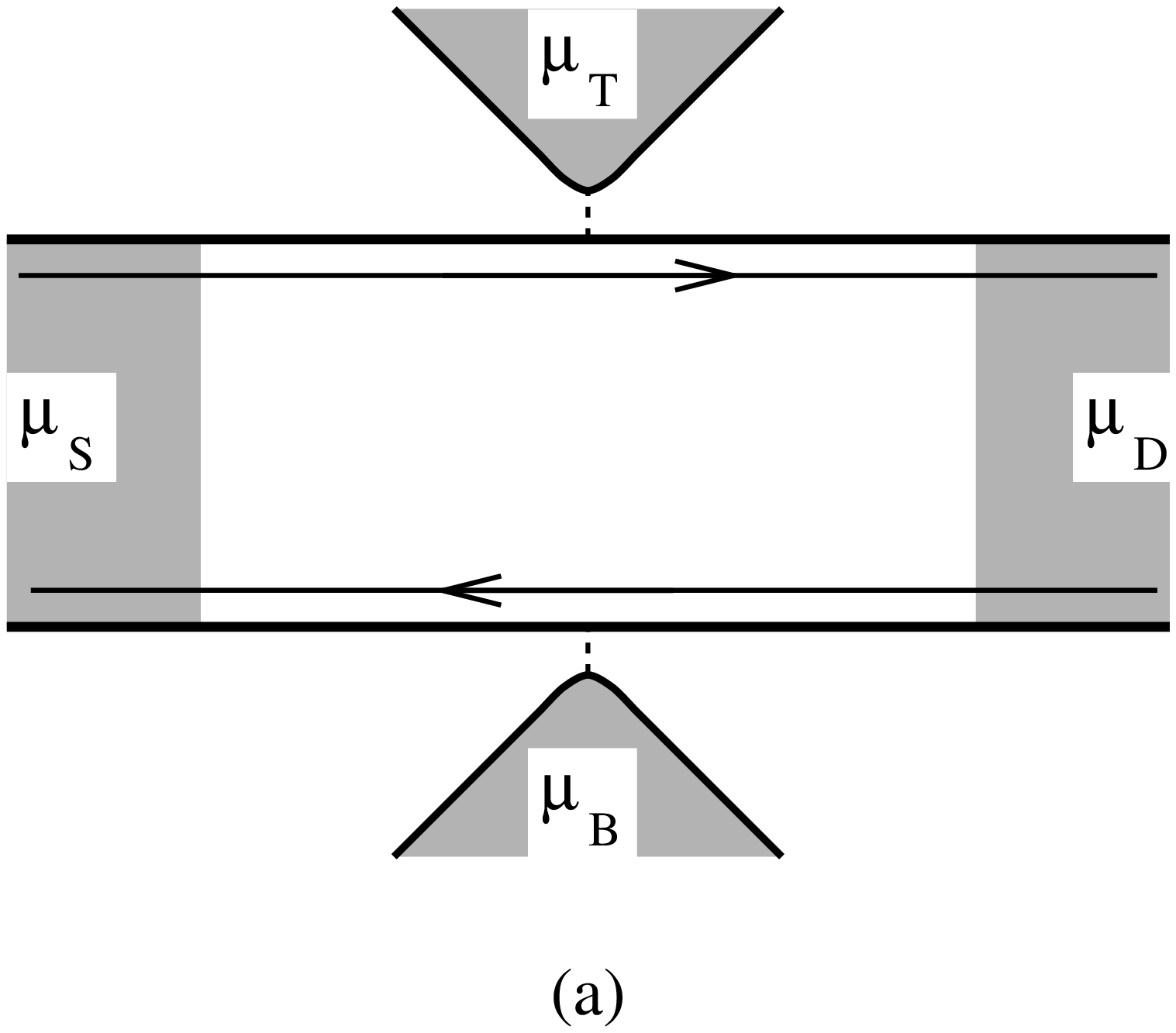,height=2.0in}}
\centerline{\psfig{figure=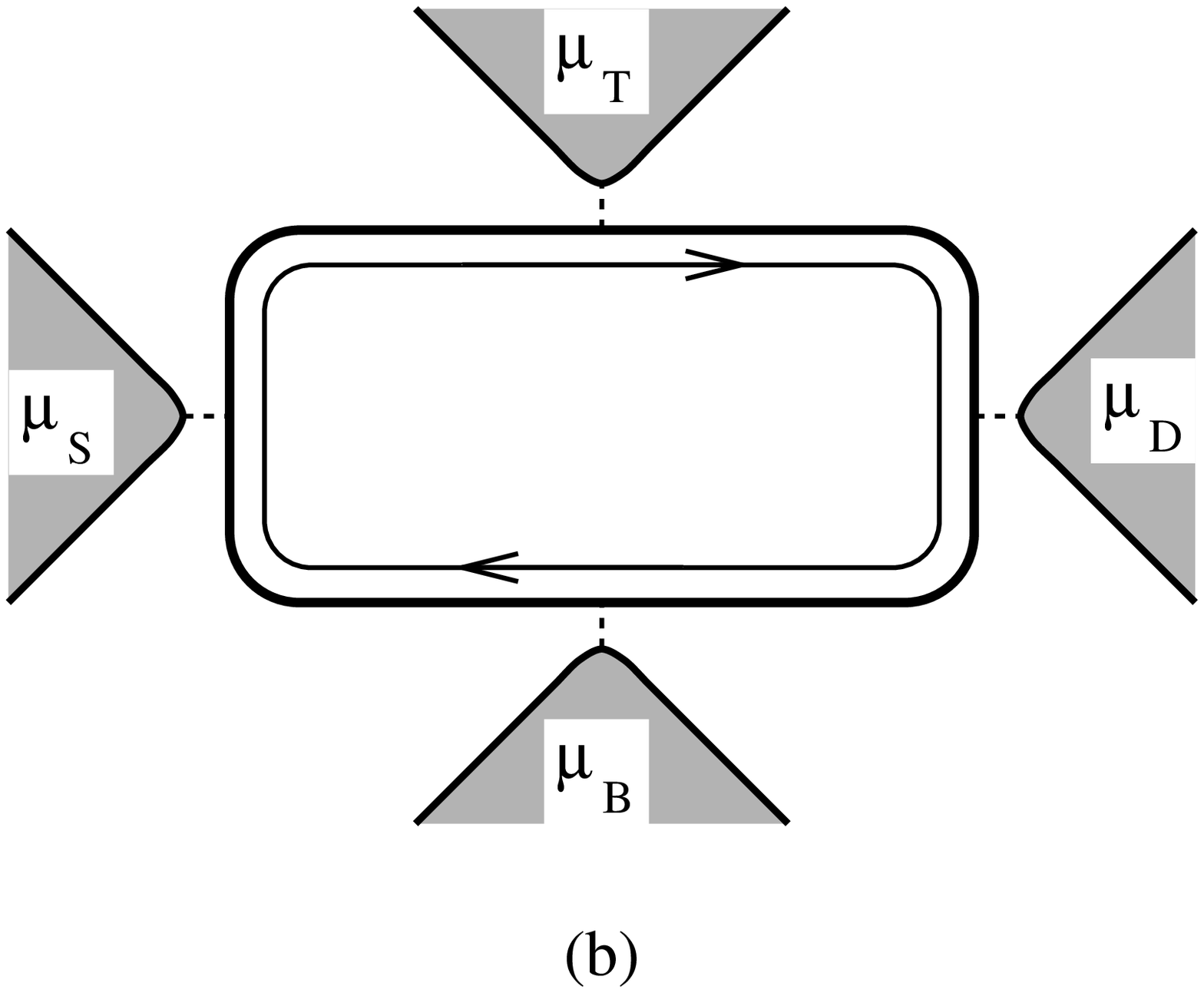,height=2.0in}}
\centerline{\psfig{figure=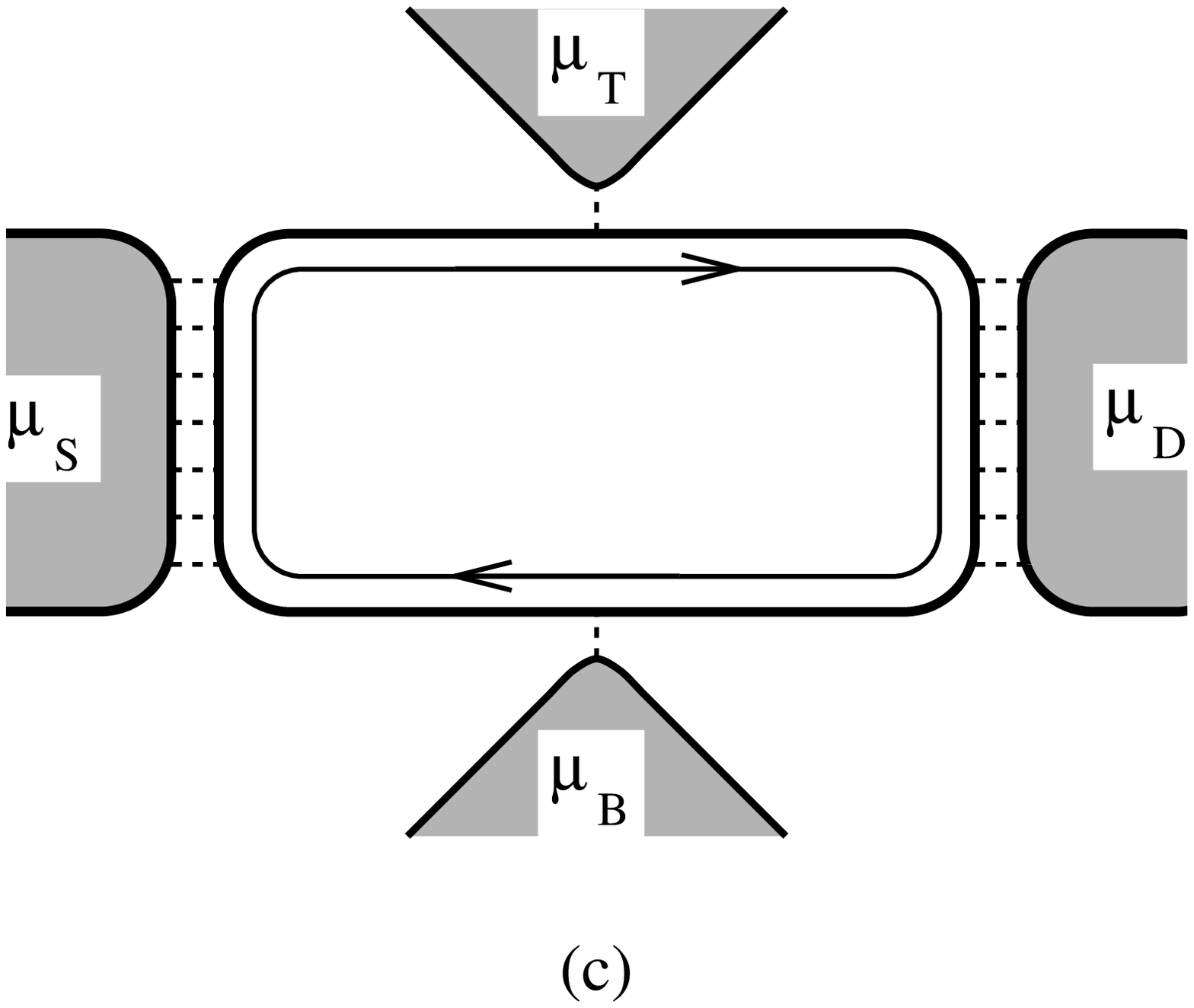,height=2.0in}}
\caption{Three different models for the source and drain
contacts to a quantum Hall fluid.  (a) 
``ideal" contacts: the FQHE edge channels
extend to infinity in both the source and drain electrodes;  (b) tunnel junction
point contacts; (c) tunnel junction line contacts. 
In (a), (b) and (c), the Hall voltage is measured via 
weakly coupled point contact tunnel junctions.
}
\end{figure}

The large resistance associated with the 
contacts will dominate the resistance in a two-terminal measurement.
To see this explicitly from the kinetic equation,
consider a two-terminal geometry (Fig. 1b),
with source electrode at $\mu_S$ and drain at $\mu_D$.
In steady state, the time derivative can be dropped, and
the kinetic equation can be readily
solved.  Denoting the current flowing along
the top and bottom edges 
by $J_{T/B}$, one obtains from the kinetic equation,
\begin{equation}
J_T - J_B = G_c \left(\mu_S - {h\over {\nu e}} J_B\right) ,
\end{equation}
\begin{equation}
J_B-J_T = G_c\left(\mu_D - {h\over {\nu e}} J_T\right)  ,
\end{equation}
where for simplicity we have assumed equal contact resistances,
$G_c^{-1}$, for the source
and drain electrodes.  The transport current, $I=J_T-J_B$
is then found to be:
\begin{equation}
I  \simeq {1\over 2}G_c(\mu_S - \mu_D)  ,
\end{equation}
under the assumption, $G_c << h/e$.  As expected,
the two-terminal resistance is simply
a sum of the two contact resistances, and is not quantized.

In a four terminal Hall measurement there are two additional voltage contacts,
on the top and bottom edges, as shown in Fig. 1b.  The chemical potentials
in these contacts, denoted $\mu_{T/B}$, are set by the requirement
that no net current flows from these electrodes into the sample.
>From (2.12) this implies 
$\mu_T = J_T h/(\nu e)$ and $\mu_B = J_B h/(\nu e)$.  The transport current,
$I=J_T-J_B$, is then,
\begin{equation}
I = {\nu e\over h}(\mu_T - \mu_B) ,
\end{equation}  
giving an appropriately quantized four-terminal Hall conductance.

As expected, we find a quantized four-terminal Hall conductance,
independent of the contact resistance.
As we show in Section III, this result breaks down
when multiple edge modes are present - 
the four-terminal conductance is {\it not}
quantized when measured with such non-ideal contacts.

\subsubsection{The Tunnel Junction Line Contact}

The above tunnel junction point contact model provides
an explicit realization of a ``non-ideal" contact,
with large contact resistance.  We now generalize this model
to describe an ``ideal contact" with vanishing contact resistance.
Consider a metallic electrode which is coupled to the edge mode
via tunneling, along a segment of length $L$. 
We refer to this as a ``line junction" contact.
The validity of a kinetic equation description again 
requires that successive tunneling events from electrode to edge
are incoherent.  This will be satisfied
provided the tunneling conductance per unit length
is sufficiently small.  Since the contact length, $L$, can be made
large, however, the total conductance between electrode and edge need
not be small.

The kinetic equation is again given by (2.11).  However,
the tunneling current from the electrode 
is now extended over a length $L$,
\begin{equation}
I_c(x) = \sigma_c \left(\mu - {h\over {\nu e}} J(x)\right) 
\ \ \ \  {\rm     for } \ \ \ \ 0<x<L
\end{equation}
where $\sigma_c$ is the tunneling conductance per unit length.
As before, $\sigma_c$ may be temperature
dependent.

In a steady state, the kinetic equation simplifies to
$\partial_x J(x) = I_c(x)$, and can be readily solved.
In the region $0<x<L$, the solution for $J(x)$ is
\begin{equation}
J(x) = {\nu e\over h} \mu + \left( J(x=0)  - {\nu e\over h} \mu \right) 
e^{-x/\ell_c} 
\end{equation}
with an equilibration length defined by: $\ell_c = \nu \sigma_c^{-1}$.
Provided $\ell_c << L$, the edge mode equilibrates fully with the
metallic electrode and $J(x>L) = \nu\mu$.   
The two-terminal conductance measured
with ``line contacts" is determined by
considering currents on the top and bottom edges.
Referring to Fig. 1c, we have $J_T = (\nu e/h) \mu_S$ and
$J_B = (\nu e/h) \mu_D$, which gives for the transport current $I = J_T - J_B$:
\begin{equation}
I = {\nu e\over h} (\mu_S - \mu_D )  .
\end{equation}
The two-terminal conductance is quantized,
indicating that the ``contact resistance" vanishes.
The ``line contact" thus provides an explicit realization
of an ``ideal contact" for FQHE states with a single edge mode.
Before discussing multiple edge modes, we briefly consider 
the role of bulk currents for $\nu^{-1}$ an odd integer.

\subsection{Bulk currents}

A confusing aspect of transport in the quantum Hall effect is
the relative importance of edge versus bulk currents. 
If the long-ranged Coulomb
interactions are screened by a ground plane,
one expects the linear response transport current to be confined to the edges.
However, when Coulomb forces are unscreened,
or one is well outside the linear response regime, 
additional bulk currents are expected, along with
bulk electric fields. 
The total transport current will be
a sum of the edge and bulk contributions.  Likewise, the measured Hall
voltage will be a sum of the edge chemical potentials and
the bulk electric potential drop. 

The distinction between edge and bulk currents becomes clear
in the IQHE for non-interacting
electrons.  Consider the schematic plot of energy levels\cite{Halperin1}
in the lowest two Landau levels, as one transverses the sample, see Fig. 2. 
In 2(a) there is no bulk electric field - the Landau levels are flat
and no bulk currents flow.
In 2(b), the presence of a bulk electric 
field gives rise to a finite velocity $v = dE/dk$ of 
the bulk states, and hence a bulk current.  In both
cases, however, the total current, summing
over both bulk and edge states, is given by 
$I =  (e/h)(\mu_T - \mu_B)$. 
\begin{figure}
\centerline{\psfig{figure=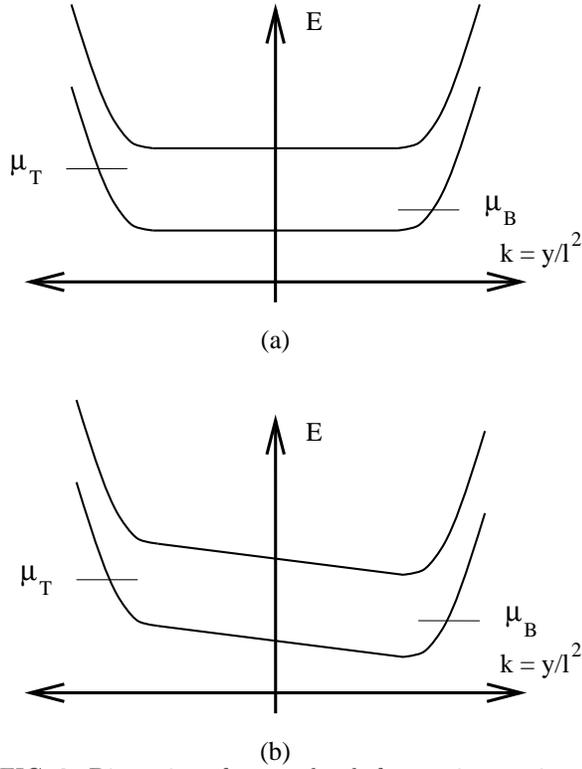,height=4.0in}}
\caption{Dispersion of energy levels for non interacting
electrons in a quantum Hall bar as a function of the
one dimensional momentum $k$, which is related to the
transverse position $y$.
In (a) there is no electric field, so the bulk states in
each Landau level are degenerate.  In (b), there is a finite
electric field in the bulk, which gives rise to a finite
velocity $v = \partial E/\partial k$ for states in the bulk.
In both cases the net current is determined by the electrochemical
potentials at the two edges, $\mu_T$ and $\mu_B$.
}
\end{figure}

This picture can readily be generalized to the FQHE.  
The total transport current in the
x-direction is expressed as a sum of bulk and edge
contributions
\begin{equation}
I = J_T - J_B 
+ \int_{y_B}^{y_T} dy \ \ J^x_{\rm bulk}  ,
\end{equation}
where the bulk current density, defined between 
$y=y_B$ and $y=y_T$ near the top and bottom edges, is
\begin{equation}
\vec J_{\rm bulk} = \sigma_{xy} \hat z \times \vec E .
\end{equation}
The edge currents on the top and bottom are denoted
$J_T$ and $J_B$.
The dividing lines, at $y_T$ and $y_B$, between the edge and bulk are
arbitrary, so that $J_{T/B}$ is only defined up to an
additive constant.  This constant may be chosen so that the
edge currents vanish in equilibrium. 
This is equivalent to normal ordering the edge current operator with
respect to the equilibrium ground state.

In the presence of bulk electric fields, $\vec{E}$, the edge current
 is no longer
conserved, since currents can flow from the edge into the bulk.
The kinetic equation obeyed by the edge currents must therefore
be modified,
\begin{equation}
\partial_t n_{T/B} + \partial_x J_{T/B} = 
\sigma_{xy} E_x (x,y_{T/B}) ,
\end{equation} 
for top and bottom respectively.

However, it is now a simple matter to eliminate the dependence
on electric fields in the above two equations.
In terms of an electric potential, $\vec E = \vec \nabla V$,
we define new currents:
\begin{equation}
\tilde J_{T/B} = J_{T/B} - \sigma_{xy} V(x,y_{T/B})  .
\end{equation}
These new currents are conserved even with bulk
electric fields present, since (2.23) can be re-written:
$\partial_t n_{T/B} + \partial_x \tilde J_{T/B}=0$.
Moreover, the total transport current becomes simply
\begin{equation}
I = \tilde J_T - \tilde J_B   .
\end{equation}
Notice that the 
new currents, $\tilde J_{T/B}$, satisfy the same steady-state
kinetic equations 
as the edge currents do in the absence of bulk electric fields.
Thus, the results of this and the next sections
are not modified by the presence of bulk electric fields.

\section{MULTIPLE EDGE MODES}

\subsection{Kinetic Equation}

For hierarchical Hall states\cite{HaldaneHalperin}
 there are multiple modes on a given edge.
The structure of the edge modes is set by the topological order
in the bulk, which is characterized by a square symmetric matrix $K$
\cite{Read,WenZee},
with integer matrix elements.
At the n'th level of the
hierarchy the matrix $K$ is an n by n matrix.
In addition there is a vector of integer ``charges",  $t_i$.
The filling factor is given by
\begin{equation}
\nu = \sum_{ij} t_i K^{-1}_{ij} t_j  .
\end{equation}
The explicit form of the $K$ matrix for a given quantum Hall state
depends on the choice of basis, as do the integers $t_i$\cite{Read,WenZee}.  
For convenience we adopt throughout the ``symmetric basis" 
in which $t_i =1$ for all $i$.

The form of the $K$-matrix determines
the structure of edge excitations. 
In terms of $n$ bosonic fields, $\phi_i$,
which satisfy commutation relations:
\begin{equation}
[\phi_i(x),\phi_j(x^\prime)] = i \pi K^{-1}_{ij} {\rm sgn}(x-x^\prime) .
\end{equation}
the appropriate edge Hamiltonian is\begin{equation}
H = \int dx {1\over {4\pi}} 
\sum_{ij} V_{ij} \partial_x \phi_i \partial_x \phi_j   .
\end{equation}
The matrix $V_{ij}$ is a non-universal positive definite matrix,
depending on the edge confining potential
and edge electron interactions.  The Hamiltonian describes
$n$ propagating chiral modes.  The directions of propagation are
determined by the signs of the eigenvalues of the $K$ matrix. 

The total electronic  
charge density at the edge is given by
\begin{equation}
n_\rho(x) = \sum_i n_i (x) 
\end{equation}
where the density in the i'th mode is,
$n_i = \partial_x  \phi_i/2\pi$.

A kinetic equation description of edge transport
follows again from the Heisenburg equations of motion,
which can be cast in the form: 
\begin{equation}
\partial_t n_i + \partial_x J_i =0,
\end{equation}
with currents:
\begin{equation}
J_i = K^{-1}_{il} V_{lj} n_j  .
\end{equation}
The expression (3.6) relating currents and densities
is analogous to the relationship
between the current and density of quasiparticles in a 
Fermi liquid.  The interaction terms $V_{ij}$ play a 
role very similar to that of the Fermi liquid parameters.

Once again, it is useful
to assume normal ordering for
the densities, $n_i$, so that in equilibrium
all densities and currents vanish.
With non-vanishing chemical
potentials, $\mu_i$, however, currents will flow.  As before,
the currents can be computed by adding to the Hamiltonian:
\begin{equation}
\delta H = - \sum_i \int dx \mu_i n_i  ,
\end{equation}
and then evaluating the currents, $J_i = -e\partial_t \phi_i/2\pi$,
using the commutation relations (3.2).  This gives
a non-vanishing transport current,
\begin{equation}
J_i = (e/h) K^{-1}_{ij} \mu_j   .
\end{equation}
When the edge modes are in equilibrium
at a common chemical potential, $\mu_i = \mu$, 
the total edge current
is appropriately quantized, as follows from (3.1):
\begin{equation}
J_\rho = \sum_i J_i = {\nu e\over h} \mu   .
\end{equation}
However, if the edge modes are fed by non-ideal contacts,
they will generally not all be at the same chemical potential.
In this case, perfect Hall quantization can break down, as
we detail below.

For simplicity we will focus hereafter on the special case of
two edge modes, where $K$ is a 2 by 2 matrix.
This includes bulk states at filling
$\nu = 1/(p_1- 1/p_2)$, with $p_1$ and $p_2$ 
odd and even integers, respectively.  
The explicit form for $K$ in the ``symmetric" basis is:
\begin{equation}
K = \left(
\begin{array}{cc}
 p_1 &  p_1-1 \\
 p_1-1  & p_1+p_2-2
\end{array}\right).
\end{equation}
When $p_1=1$ the $K$ matrix is diagonal, with
eigenvalues $1$ and $p_2-1$. 
When 
$p_2$ is positive, as for $\nu=4/3$ and $\nu=2$,
both modes propagate in the same direction.
For negative $p_2$, however,
the two modes are predicted to propagate in opposite
directions.  This includes fillings $\nu=2/3$ and $\nu=4/5$.

As we shall see below, quantization of the Hall conductance
when multiple modes are present, generally requires
that the different modes on a given edge
are equilibrated at a common chemical potential.
For this reason it will be convenient to 
transform to a new set of fields which reveal
more readily when an edge is equilibrated.
To this end we define charge and neutral fields\cite{KFP}
via
\begin{eqnarray}
\phi_\rho &&= \phi_1 + \phi_2  ,\\
\phi_\sigma &&= \phi_1 + (1- p_2) \phi_2.
\end{eqnarray}
The charge and neutral fields commute with one another,
and satisfy:
\begin{equation}
[\phi_\rho(x), \phi_\rho(x^\prime)] = i \pi \nu {\rm sgn}(x-x^\prime) ,
\end{equation}
\begin{equation}
[\phi_\sigma(x), \phi_\sigma(x^\prime)] = i \pi p_2 {\rm sgn}(x-x^\prime) .
\end{equation}
Note that $p_2$ can be negative.
The Hamiltonian becomes,
$H = H_\rho + H_\sigma + H_{\rm int}$, 
with charge and neutral pieces
\begin{equation}
H_\rho =  {v_\rho\over {4\pi\nu}}  \int dx (\partial_x\phi_\rho)^2
\end{equation}
\begin{equation}
H_\sigma = {v_\sigma \over {4\pi|p_2|}}  \int dx 
( \partial_x\phi_\sigma)^2
\end{equation}
coupled together via
\begin{equation}
H_{\rm int} = {2 v_{\rm int}\over{4\pi}}\int dx \partial_x\phi_\rho
\partial_x\phi_\sigma  .
\end{equation}
The velocities $v_\rho$, $v_\sigma$ and $v_{\rm int}$ depend on the
original velocities, $V_{ij}$ in (3.3).

The equations of motion for
the charge and neutral fields can be obtained
and expressed in terms of charge and neutral densities:
\begin{equation}
n_\alpha = {1\over 2 \pi} \partial_x \phi_\alpha
\end{equation}
with $\alpha = \rho, \sigma$.  Again, they can be written
as continuity equations,
\begin{equation}
\partial_t n_\alpha + \partial_x J_\alpha = 0 
\end{equation}
with currents defined as 
\begin{equation}
J_\rho = v_\rho n_\rho + \nu v_{int} n_\sigma ,
\end{equation}
and
\begin{equation}
J_\sigma = {\rm sgn}(p_2) v_\sigma  n_\sigma + p_2 v_{int} n_\rho  .
\end{equation}
Notice that for $p_2 <0$, the neutral mode propogates in the direction
opposite to the charge mode.

The charge and neutral currents can be given a simple physical
interpretation.  Using (3.11) the charge current can
be expressed in terms of the original currents, $J_i$, as
\begin{equation}
J_\rho = J_1 + J_2  .
\end{equation}
Thus $J_\rho$ is simply the total electrical charge current along
the edge.  Likewise, the neutral current takes the form  
\begin{equation}
J_\sigma = J_1 + (1-p_2) J_2  .
\end{equation}
In the presence of non-vanishing 
chemical potentials, $\mu_i$, this can be re-expressed using (3.8) as:
\begin{equation}
J_\sigma = {e\over h} (\mu_1 - \mu_2).
\end{equation}
When the edge is equilibrated, $\mu_1=\mu_2$, the neutral current vanishes,
whereas a non-zero $J_\sigma$ indicates
an un-equilibrated edge.
Thus $J_\sigma$ can be interpreted
as a current of (neutral) vortices, moving along the edge.
The flux of vortices leads to a chemical potential gradient
between the two edge modes.

\subsection{Contacts}

In this sub-section we consider models for contacts appropriate
to quantum Hall fluids with multiple edge modes.

\subsubsection{The Tunnel Junction Point contact}

We first consider leads which are connected to the sample via
tunnel junction point 
contacts.  When multiple edge channels are present, the
electrons from the leads 
can tunnel onto the edge in
different ways.  For instance, for $\nu=2$ an electron 
can tunnel into either of the two edge modes.
Generally, the tunneling rates will be different, depending on the details
of the tunnel junction.
As a result, 
the two edge modes will be populated differently,
at different chemical potentials.
As we shall show, this leads to a breakdown in the
quantized Hall conductance (provided other processes
do not equilibrate the modes - see below).
This can be seen easily in the extreme limit
that the tunneling is only into the outer edge
channel, in which case 
the Hall conductance would be 
$1 e^2/h$ rather than $2 e^2/h$.

Consider then a metallic electrode
at chemical potential $\mu$ connected to 
the two-channel edge via a tunnel junction point contact.  
In addition to the transfer of an electron to one of the two edge modes,
the tunneling process
may involve the simultaneous transfer of other electrons between the 
two modes.
The most general charge $Q=1$ process 
adds an electron to one channel, say channel one, and  
transfers $m$ electrons from channel two to one.  Here $m$ is
an arbitrary integer.
This process is equivalent to adding a unit charge to 
the charge mode, $\phi_\rho$, while creating an
instanton of amplitude $(2\pi)(1 + m p_2)$ in the neutral field,
$\phi_\sigma$.  This instanton can be interpreted as the addition of
$(1+mp_2)$ vortices.  
Upon using the commutation relations (3.13)-(3.14), one can show that this  
combined process may
be accomplished via the operator $\exp i\phi_m(x_i)$ with
\begin{equation}
\phi_m = {1\over \nu} \phi_\rho + ({1\over p_2} + m) \phi_\sigma  .
\end{equation}
Electron charge transfer between the lead and the quantum Hall
edge may then be introduced via a tunneling term in the
Hamiltonian,
\begin{equation}
H_{\rm tun.} =  - \sum_m t_{m} \psi(x=0) \int dx \delta(x)
e^{i \phi_m(x) }  + c.c..
\end{equation}
As in the single channel case, 
the tunneling current from lead to edge may be expressed in terms of the chemical potential drop.  However, in this case
the different
channels may be at different chemical potentials.
Consider the set of chemical potentials, $\mu_m$, defined as the change
in energy when a particle is added using the m'th
tunneling operator.  These can be related to the chemical potentials
of the original two channels as $\mu_m = \mu_1 + m(\mu_1-\mu_2)$.
Finally, upon using (3.8) and (3.22-3.23), these can be re-expressed in terms of the currents as
\begin{equation}
\mu_m = {h\over e} \left( {1\over \nu} J_\rho + ({1\over p_2} + m) J_\sigma \right).
\end{equation}
The current tunneling from the lead to the edge in the m'th tunneling channel is then,
\begin{equation} 
I_m(x) = \delta(x) \tilde G_m ( \mu - \mu_m),
\end{equation}
with $\mu$ the chemical potential of the metallic electrode.
Again, the tunneling conductances $G_m$ will in general
be temperature dependent.  At low temperatures, they
are expected to vanish as a power of temperature,
$\tilde G_m \approx t_m^2 T^{\Delta_m}$.  For very low temperatures,
it is possible that a single channel (with the smallest $\Delta_m$)
will dominate the tunneling.

We now modify the kinetic equations (3.19) to include tunneling
of charge from the leads to the edge, by writing
\begin{equation}
\partial_t n_\rho + \partial_x J_\rho = I_\rho  ,
\end{equation}
\begin{equation}
\partial_t n_\sigma+ \partial_x J_\sigma =  I_\sigma  .
\end{equation}
Here $I_\rho$ and
$I_\sigma$ denote the total  
tunneling rates for charge and vorticity,
expressed as a sum over contributions from each of the
tunneling channels,
\begin{equation}
I_\rho = \sum_m  I_m  ,
\end{equation} 
\begin{equation}
I_\sigma = \sum_m (1 + m p_2) I_m.
\end{equation}

The tunneling currents from lead to edge may now be re-expressed in terms
of the edge currents themselves, $J_\rho$ and $J_\sigma$,
upon using (3.27)-(3.28).  This gives 
\begin{equation}
I_\rho = G_1 \left( {h\over{\nu e}}J_\rho(x) - \mu \right) 
          + G_2 {h\over p_2 e} J_\sigma(x)  ,
\end{equation}
and 
\begin{equation}
I_\sigma = G_2 \left( {h\over{\nu e}} J_\rho(x) - \mu\right)
            + G_3 {h\over {p_2 e}} J_\sigma(x)  ,
\end{equation}
where we have defined three conductances
\begin{equation}
G_a = \sum_{m}  \tilde G_{m} (1 + m p_2)^{a-1}
\end{equation}
with $a  = 1,2,3$.
The conductances $G_a$ give a complete characterization of the
tunnel junction between the lead and the quantum Hall edge.
In general, $G_1$, $G_2$ and $G_3$, will be of comparable
magnitudes.  As in Section IIB we will assume that
$G_a << e^2/h$.  Since $\tilde G_m > 0$, $G_1$ and $G_3$ are
necessarily nonzero and positive.  In contrast, $G_2$ can
be positive or negative.  Although a generic contact will have non-zero
$G_2$, it is possible to imagine fine tuning a contact to make
$G_2$ vanish.  

Given the conductances $G_a$ characterizing each 
contact, equations (3.29,3.30) and (3.33,3.34) can be used to 
determine transport properties for multi-terminal measurements.
Again referring to Fig. 1, let $J_{T,B \rho,\sigma}$ denote
the charge and neutral currents on the top and bottom edges.
Consider first the two terminal measurement
shown in Fig. 1b, with identical tunnel junctions connecting the sample
to the source and drain electrodes,
$G_{aS} = G_{aD}$.
Solving the steady state kinetic equations relates 
the net transport current, $I = J_{T\rho} - J_{B\rho}$,
to the chemical potentials of the source and drain electrodes,
\begin{equation}
I = {G_{1S}\over 2} (\mu_S - \mu_D).
\end {equation}
As in the single channel case, the two terminal resistance is
dominated by the contacts, equaling the sum of the two contact resistances,
$G_1^{-1}$.  

Under the above transport conditions, in addition to the flow of electrical current throught the sample, there is also a flow of
vortices - that is a non-vanishing neutral current $J_\sigma$ - 
given by
\begin{equation}
J_{T\sigma} - J_{B\sigma} = 
{G_{2S}\over 2} (\mu_S - \mu_D) = {G_{2S}\over G_{1S}} I.
\end{equation}
Note that this vortex current is proportional to $G_2$, and will 
generically 
be non-zero, unless $G_2$ is fine-tuned to zero.
Since the neutral current is proportional to the 
difference of the chemical potentials of the two edge modes, (3.24),
a non-vanishing neutral current indicates an absence of
edge equilibration.

Next consider a four terminal measurement, in which tunnel junction
contacts are also used as voltage probes
on the top and bottom edges of the sample, see Fig. 1b.  For simplicity 
we assume that $G_{aT} = G_{aB} << G_{aS} = G_{aD}$.  The 
chemical potentials of the voltage probes
$\mu_T$, $\mu_B$ are adjusted so that 
no net current flows through the contacts, $I_{\rho T} = 
I_{\rho B} =0$.  Upon using the steady state kinetic equations 
for this four-terminal geometry one finds
\begin{equation}
\mu_T - \mu_B = {h\over{\nu e}} I + {h G_{2T}\over {p_2 e G_{1T}}} 
\left( J_{T\sigma} -  J_{B\sigma}\right) ,
\end{equation}
where $I$ is the source-to-drain transport current.
The  four-terminal Hall resistance, $R_H = (\mu_T-\mu_B)/(eI)$, is given by,
\begin{equation}
R_H = {h\over e^2} \left( {1\over \nu} + 
{G_{2T} G_{2S}\over{p_2 G_{1T} G_{1S}}} \right)  .
\end{equation}
Again, unless $G_2 = 0$, the Hall resistance $R_H$ is not
quantized.  Since the two edge channels are out
of equilibrium, the sample edge does not have
a well defined chemical potential.
The voltage probes measure a weighted
average of the chemical potentials, with
the relative weights (from $G_a$) depending on 
non-universal details of the contacts.
Other more complicated multi-terminal geometries can also be
easily analyzed using the kinetic equations.

\subsubsection{The Ideal Contact}

As we have seen above, for an edge with two modes, neither the two nor four terminal
conductances is quantized when measured with tunnel junction
point contacts.
In Section II we considered an ``ideal contact", which gave a quantized conductance for both two and four terminal measurements, in the
case of a Hall fluid with a single edge mode.
Do these conclusions remain valid
for a Hall fluid with multiple edge branches?
We now show that they do, {\it provided} all of the
channels on a given edge propagate in the same direction.
For $\nu = 1/(p_1 + 1/p_2)$, this is the case for $p_2>0$
(e.g. $\nu = 2,2/5,4/3,...$).  However,
for $p_2<0$ (e.g. $\nu = 2/3,4/5,...$), when the two modes
propagate in opposite directions, we will show
that the two and four terminal conductances are {\it not}
quantized, even when measured with such ``ideal contacts".

By definition, all edge modes which emanate from an ``ideal contact" 
are in equilibrium at the reservoir chemical potential.
When both channels move in the same direction,
they will then share a common chemical potential,
having emanated from the same reservoir.  (The
neutral current, $J_\sigma$ in (3.24), will everywhere vanish.)
It then follows from (3.9) that the net transport current along the edge
will be appropriately quantized, as will the two and four terminal
conductances.
 
In contrast, when the two edge modes move in opposite directions, they will
generally be at different chemical potentials in a transport measurement,
having emanated from different ``ideal contacts". 
The two edge modes will not be in common equilibrium,
and there will be a flow of vorticity along the edge: $J_\sigma \ne 0$. 
In Ref. \onlinecite{sun} we showed that the two terminal conductance
measured with such ``ideal contacts" is given by
\begin{equation}
G = (g_+ + g_-){e^2\over h}  .
\end{equation}
Here $g_\pm$ are right (left) conductances,
defined as the change in current
in response to a chemical potential which couples only
to the right (left) moving modes.  Both conductances
$g_\pm$ are positive and satisfy $g_+ - g_- = \nu$,
but they are non universal and depend on the interaction matrix
$V_{ij}$ in (3.3).   It follows that $G > \nu e^2/h$ is nonuniversal.
One can likewise show that a four terminal conductance measured 
using four ``ideal contacts" is also non-universal.

In Section II
we demonstrated that
an ``ideal contact" could be realized
more microscopically as a tunnel junction ``line contact".
The ``line contact" junction also eliminated the need for fractional edge modes to exist 
inside the reservoirs. 
Does the ``line junction", when generalized to an
edge with multiple modes, restore universality
absent with the ``ideal contacts"?
We now show that this is { \it not} the case.

\subsubsection{The Line Junction Contact}

Consider then a tunnel junction 
``line contact'' coupling to an edge with two modes.
The line contact may be characterized by three conductivitities
$\sigma_a$, ($a=1,2,3$),
defined as 
the tunneling conductances, $G_a$ in (3.35),  per unit length.  
For a contact with length $L$ the tunneling currents from lead to the edge
can then be expressed as,
\begin{equation}
I_\rho(x)  = \sigma_1 \left( {h \over { \nu e}}J_\rho(x) - \mu \right) 
          + \sigma_2 {h\over {p_2 e}} J_\sigma(x)  ,
\end{equation}
\begin{equation}
I_\sigma(x) = \sigma_2 \left( {h\over {\nu e}} J_\rho(x) - \mu\right)
            + \sigma_3 {h\over {p_2 e}} J_\sigma(x)  ,
\end{equation}
for $0<x<L$.  The ``line contact" is modeled
by adding these spatially dependent source terms,
to the right hand side of the kinetic equations
(3.29)-(3.30).

In the steady state the kinetic equations can be readily solved
by diagonalizing a 2 by 2 matrix for the currents $J_\rho$ and $J_\sigma$.
In the region of the line contact $0<x<L$ the solution
takes the form,
\begin{equation}
\left( \begin{array}{c} J_\rho \\ J_\sigma  \end{array} \right)
= {\nu e \over h}\mu \left( \begin{array}{c} 1 \\ 0  \end{array} \right)
+ c_1 e^{x/\ell_1} 
\left( \begin{array}{c} a_1 \\ b_1  \end{array} \right)
+ c_2 e^{x/\ell_2} 
\left( \begin{array}{c} a_2 \\ b_2  \end{array} \right)
\end{equation}
where $\ell_{1,2}^{-1}$ are eigenvalues of the matrix and $a_{1,2},b_{1,2}$ are its eigenvectors.  Here $\mu$ is the chemical potential
of the contact.
The constants $c_1$ and $c_2$ are determined by the boundary
conditions at $x=0,L$.  

Using (3.35) and the explicit formula for the eigenvalues, it
can be shown that ${\rm sgn} \ell_1^{-1} \ell_2^{-1} = {\rm sgn} p_2$.
Thus, when $p_2>0$ and both edge modes move in the same direction,
both solutions decay exponentially.  Then provided $L>>\ell_1,\ell_2$,
the edge modes emanating from the ``line contact" will be
fully equilibrated with the contact: $J_\rho = (\nu e/h) \mu$ ,
$J_\sigma = 0$.  It then follows that all measured conductances
will be appropriately quantized.

However, when $p_2<0$, one solution in (3.43) is growing exponentially,
while the other is decaying.
Then generically the neutral current $J_\sigma$ will be non-zero
at the endpoints of the ``line contact".  Again,
the presence of a non-vanishing neutral current indicates that
the two edge modes are not in equilibrium with one another.
This in turn implies a non-universal Hall conductance,
for both two and four terminal measurements, just as for
the ``ideal contact" model.
However, in contrast to the ``ideal contact",
the value of the non quantized conductance is determined by 
the ratios of the tunneling conductances $\sigma_a$ and is
independent of the nonuniversal interaction matrix $V_{ij}$.

So far, all the models of contacts that we have considered
lead to an absence of conductance quantization for an edge with
two modes moving in opposite directions.  The lack of quantization is due to
an absence of equilibration between the oppositely moving modes. 
Real quantum Hall samples show precise quantization, presumably due to
processes along the sample edges which allow for equilibration.
We now turn to a discussion of impurity scattering along the edge,
and show how it equilibrates and restores quantization.

\subsection{Edge Equilibration: Random impurities}

Consider impurity scattering at the edge which allows
for non-momentum conserving
charge transfer processes between nearby edge modes.
It is useful to distinguish two length scales.
The first,  
a tunneling mean free path $\ell$, denotes the distance an electron
propagates along the edge before it is scattered by
an impurity into a different
channel.  For non interacting electrons, this scattering is elastic,
and $\ell$ is temperature independent.  For fractional quantum Hall
edge channels, though, this length can be temperature dependent,
and even divergent at zero temperature (see below).
A second length, denoted $\ell_\phi$, is the length
over which electrons lose their phase coherence within a single edge channel.
In general,
the dephasing length diverges at low temperatures.  It arises
both due to thermal dephasing (which gives 
$\ell_\phi \approx v/k_B T$) and due to inelastic scattering off phonons
or other electrons, 
for which $\ell_\phi$ diverges as a different power of the temperature.

In order to measure a quantized Hall conductance,
the separation $L$ between current and voltage leads must exceed
both $\ell_\phi$ and $\ell$.
On scales beyond $\ell$ one expects the multiple modes to have equilibrated.
However, $L$ must also be larger than $\ell_\phi$
for robust quantization, since in the
regime $\ell_\phi>L>\ell$, sample specific
mesoscopic fluctuations in the measured conductance are expected.
True equilibrium is thus reached at the
larger of $\ell$ and $\ell_\phi$.  

We now describe the equilibration
due to random impurities, which gives a finite length $\ell$
for inter-channel mixing.  
The operator which transfers a unit charge between
the two edge modes is $\exp i\phi_\sigma$, as can be deduced from the definitions in
IIIa.  
Edge impurity scattering
can thus be incorporated into (3.15)-(3.17) by adding a term to the
Hamiltonian of the form\cite{KFP,sun},
\begin{equation}
H_{\rm random} = \int dx \left[
\xi(x)e^{i\phi_\sigma} + c.c. \right],
\end{equation}
where $\xi(x)$ is a (complex) spatially random tunneling amplitude.
Since the total charge is conserved in such a scattering event,
the kinetic equation for the charge sector is unchanged:
\begin{equation}
\partial_t n_\rho + \partial_x J_\rho = 0  .
\end{equation}
However, when an electron tunnels between the two channels,
$p_2$ vortices are destroyed.
The neutral sector thus contains a ``collision term", 
$I_\perp(x)$.
\begin{equation}
\partial_t n_\sigma + \partial_x J_\sigma  =  I_{\perp}(x).
\end{equation}

We compute the tunneling current, $I_\perp(x)$, using Fermi's golden rule.  
Again, such a description is appropriate in the limit that successive
intermode tunneling events are incoherent.
Provided the chemical potential difference, $J_\sigma = (e/h)(\mu_1 - \mu_2)$,
between the two
edge modes is sufficiently small, the response will be Ohmic, and
the tunneling current per unit length will be linear in $J_\sigma$,
\begin{equation}
I_{\perp} = - {\rm sgn}( p_2 ) J_\sigma/\ell  .
\end{equation}
Here $\ell$ is the inter-mode scattering length.

For the simple case of disorder with a
delta-function correlated gaussian distribution,
the length $\ell$ may explicitly be computed perturbatively
in the variance $W = < |\xi(x)|^2 >$.
We find,
\begin{equation}
\ell ^{-1} \sim \ell_0^{-1} ({T\over T_0})^\alpha,
\end{equation}
where $\ell_0^{-1} \propto W$ and $T_0$ is the high energy
cutoff.  The exponent $\alpha$ is related to 
the scaling dimension $\Delta$ of the tunneling operator 
$\exp i\phi_\sigma$ via $\alpha = 2\Delta - 2$\cite{KFP,sun}.
For positive $p_2$ one has $\Delta = p_2/2$, so that
$\alpha = p_2 - 2$.  For negative $p_2$, however,
$\Delta \ge |p_2|/2$ is non-universal and depends on the
coupling constants $V_{ij}$ in the Hamiltonian (3.3).  In this
case, $\alpha \ge |p_2| -2$.

In the steady state, the extra term 
on the right hand side of (3.46) causes an
exponential decay
of the neutral mode with length $\ell$,
\begin{equation}
J_\sigma(x) = J_\sigma(0) e^{- {\rm sgn}(p_2) x /\ell}. 
\end{equation}
Beyond this scale,
the neutral current vanishes.  Since the neutral current
is proportional to the difference between chemical
potentials of the two modes, (3.24), this indicates that
the edge equilibrates on this length scale.
When the spacing between current and voltage probes is larger than
$\ell$, the conductance will be appropriately quantized 
regardless of the nature of the contacts.

Recall that for $p_2<0$, the neutral mode propagates
{\it upstream}, in the direction opposite to the charge mode, $J_\rho$.
It is interesting to note that the sign of $p_2$ also determines the
direction in which the neutral mode decays.  When $p_2<0$, 
$J_\sigma$ likewise decays in the  upstream direction.
It then follows that when the contacts are further than
$\ell$ apart, the departure from equilibrium will only 
be present within a distance $\ell$ upstream from the
contacts.

Since Fermi's Golden rule which lead to the result (3.47)
can only be trusted when successive tunneling events are uncorrelated,  
these equations must be
used with caution at very low tempertures, when
quantum coherence effects can become important.
There are two possible low temperature regimes,
depending on the value of the exponent $\Delta$. 

For $\Delta < 3/2$, weak disorder is relevant,
and the renormalization group flows are to a disorder dominated
fixed point with $\Delta = 1$ at the fixed point.
In this case, the above kinetic equation,
which predicts only a single propagating mode, is not correct.
Although the decay of the neutral mode
$J_\sigma$ on scale $\ell$ is correct, there is a hidden
(neutral) mode which propagates at zero temperature.  
This hidden mode is not a vortex current, and can
be present even on a fully equilibrated edge
($J_\sigma = \mu_1 - \mu_2 =0$).  As shown in Ref. \onlinecite{sun},
this hidden mode itself decays away at finite temperatures. 
Provided we focus on those properties which are insensitive
to the presence of this hidden mode, the kinetic equation 
provides an adequate description even in this regime.

For $\Delta > 3/2$ the length $\ell$ diverges 
in the zero temperature limit, and at $T=0$ both modes
propagate.   In this case, the kinetic equation gives a correct
description even as $T \rightarrow 0$.
This is plausible on physical grounds, since in this case
$\ell(T)$ diverges more rapidly than the phase breaking length,
$\ell_\phi \sim v/k_BT$, so that successive tunneling events should
be incoherent.  On a more technical level,
the validity of (3.48) follows from the  
perturbative irrelevance of $W$ in the renormalization
group calculation.

The exponential decay of the neutral mode described above 
is a result of the linear (or Ohmic) inter-mode tunneling current in
 response to
a non equilibrium chemical potential difference between the
two modes.  Even when this Ohmic response vanishes at $T=0$
 (for $\Delta \ge 3/2$),
a finite {\it non Ohmic} tunneling current is expected,
which will vanish faster than linearly with
the chemical potential difference.  This
non linear tunneling current can also lead to equilibration
between the edge channels.  But in this case,
the decay of $J_\sigma$ is not exponential as in (3.49), but
rather algebraic.  Specifically, the decay of $J_\sigma$ at $T=0$ can 
be studied by analyzing a {\it nonlinear} kinetic equation.
Consider the steady state version of (3.46) with a
non-linear tunneling current, 
\begin{equation}
I_\perp \propto \ell_0^{-1}  
 ({J_\sigma\over J_0})^\alpha J_\sigma  ,
\end{equation}
where $J_0 = k_B T_0 e/h$.
Then (3.46) may be readily integrated to give
\begin{equation}
J_\sigma(x) = {J_\sigma(0) \over {\left[ 1 +
{\alpha x\over \ell_0} (J_\sigma(0) / J_0)^{\alpha}\right]^{1/\alpha}}}  .
\end{equation}
At large distances, $J_\sigma(x)$ decays algebraically, varying 
as $J_0 (\alpha x/\ell_0)^{-1/\alpha}$, independent of $J_\sigma(0)$.
In section IV we will explore the consequences of this slow decay
for quantum Hall states at filling $\nu = 4/3$ and $4/5$.

\subsection{The Compound Contact}

As we have seen, the absence of a quantized conductance
can be traced to an absence of equilibration between the multiple
edge modes.  However, when impurity scattering is included along the edges,
the different modes will equilibrate on a scale set by $\ell$.
Then, provided the spacing between the current and voltage contacts is larger than $\ell$, an appropriately quantized conducance will be found
regardless of the nature of the contacts.

For a perfectly clean system with vanishing impurity scattering at the edge,
one might ask whether it is still possible to construct a contact
which populates all the edge modes in equilibrium?  
Such a {\it fully equilibrated} contact may be constructed,
as illustrated in Fig. 3.
Consider a ``compound" contact which consists of a tunnel junction point
contact, say, with edge impurity scattering regions localized within a length
scale $L$ on either side\cite{thanks}.  Provided $L$ is larger than $\ell$,
all of the currents away from the compound contact will
then be fully equilibrated.  Outside the contact
region, there will be a unique edge chemical
potential, or equivalently a vanishing neutral current density.
The key point which distinguishes this ``compound contact" from the earlier 
models for contacts, is that it allows for direct transfer of charges
between the two oppositely moving edge modes.  In this way, the
current in the right moving mode, can lead to an adjustment of the
left moving current, causing equilibration.
It should be emphasized, though, that in some cases (eg. $\nu=4/5$)
the length scale $\ell$ diverges at zero temperature, so such a compound
contact would eventually cease to equilibrate at low enough temperatures.

\begin{figure}
\centerline{\psfig{figure=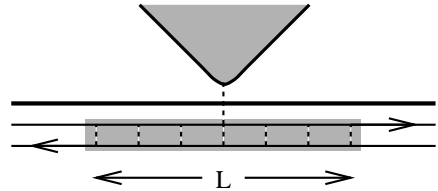,height=1.0in}}
\caption{
A compound contact.  Edge impurity scattering
allows for equilibration between the different edge channels within
a distance $L$ of the contact, as indicated by the shaded region.
}
\end{figure}

A four terminal Hall conductance measured with such
``compound" contacts will be appropriately quantized, when $L>> \ell$.
Within the scattering region of size $L$ at the compound current contacts there may be
a nonequilibrium current distribution, with $J_\sigma\ne 0$.  
However, they decay away in a distance $\ell$,
and at the voltage contacts $J_\sigma = 0$.  The voltage probes will
then measure
$\mu = J_\rho / \nu$, giving a
quantized Hall conductance.

One may also consider a compound contact made from a tunnel junction
line contact.  In this case, solving the kinetic equation 
shows that when $L >> \ell$, the edge is fully equilibrated
at the chemical potential of the ``upstream" lead.   (Here
``upstream" is defined in terms of the direction of propagation
of the charge). 
Thus, the compound line junction contact is an explicit
realization of the B\"uttiker ideal contact even when
$p_2<0$.  The two terminal conductance measured with
such contacts should be appropriately quantized.

\section{Experimental Implications: 
Lack of Equlibration for $\nu = 4/5, 4/3$}

In this section we focus on specific predictions regarding
edge state equilibration in the FQHE
which would be interesting to test experimentally.

In section IIIC we found for filling fractions with
$|p_2| \ge 4$ (such as $\nu = 4/5$ or $4/3$)
that the edge state equilibration length diverges
at low temperatures.   This should lead to a 
breakdown of the quantization of the Hall conductance
as the temperature is lowered.

\begin{figure}
\centerline{\psfig{figure=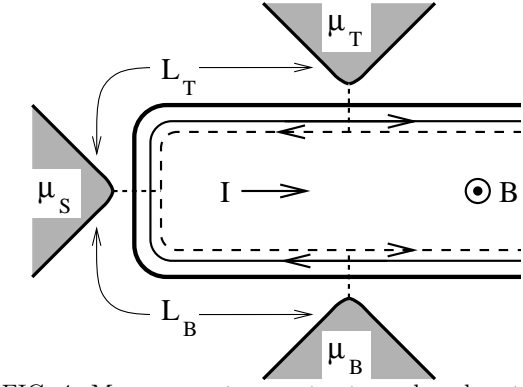,height=2.0in}}
\caption{Measurement geometry to probe edge state
equilibration.  Current is injected through the
source contact, and the Hall voltage is measured
between the contacts $T$ and $B$.  As indicated
by the solid line, charge propagates counterclockwise
along the edge.  For $p_2<0$, the neutral mode propagates
(and decays)
in the opposite direction, as indicated by the dashed line.
For $p_2 >0$ the neutral mode propagates clockwise, and the
arrow on the dashed line should be reversed.
}
\end{figure}

In order to probe the equilibration between edge states,
consider the sample geometry shown in Fig. 4.  Current
is injected through a source contact, and the Hall voltage
is measured using voltage probes on the top and bottom
edges at distances $L_T$ and $L_B$ from the source.
We assume that the drain contact is much further 
away, so that it has no effect on the 
equilibrium of the edge channels near the voltage probes.
For simplicity, as in section 3B1, we model the contacts as tunnel
junction point contacts with conductances $G_{Ta}, G_{Ba} \ll G_{Sa}$.
Further, suppose that the magnetic field points out of the
page in Fig. 4, so that that the direction of propagation
of the charge $J_\rho$ around the edge is counter-clockwise.

Since the source contact is non ideal, it populates the edge modes
out of equilibrium.  
For $\nu = 4/3$ ($p_2>0$), all of the channels propagate counter clockwise,
so that only the top edge is out of equilibrium.  
The lack of equilibrium on the top edge, characterized by
$J_{T\sigma}$, is determined by $I$ and
decays away in an equilibration length $\ell$.
Following the analysis of section 3B1, we then find that
just above the source lead $J_{T\sigma}(0) = (G_{2S}/G_{1S})I$.
It then follows that 
the measured Hall voltage is not quantized, and is given by
\begin{equation}
\mu_T - \mu_B  = {h\over e} \left( {1\over \nu} +
{G_{2T} G_{2S}\over{p_2 G_{1T} G_{1S}}} e^{-L_T / \ell}  \right) I  ,
\end{equation}
where $\ell$ is the temperature dependent equilibration length
(3.48).
Since $G_2$ in (4.1) can be either positive or negative, even the 
sign of the deviation from quantization is non universal.
Note that since the bottom edge is fully equilibrated, the 
nonuniversal linear Hall voltage
is {\it independent} of the position $L_B$ of the bottom contact.

When $\nu = 4/5$ ($p_2 < 0$), the neutral mode, 
$J_\sigma$ propagates clockwise,
in the opposite direction as $J_\rho$.  It follows that 
in this case the bottom edge is out of equilibrium, wheras
the top edge is fully equilibrated.  The measured linear Hall voltage
is given by (4.1) with T replaced by B.  Surprisingly, even
though the current is injected onto the top edge, the nonuniversal
linear Hall voltage depends only on the position 
$L_B$ of the bottom contact.

In order to estimate
the temperature at which the lack of equilibration should
be detectable, let us consider the
temperature scale $T^*$ at which the equilibration
length is comparable to the distance $L$ between contacts.  
>From (3.48),
\begin{equation}
T^* = T_0 \left({\ell_0\over L}\right)^{1/\alpha}.
\end{equation}
As a rough estimate we take
the cutoff energy $T_0$ to be equal
the 
excitation gap for the bulk Hall fluid, $T_0 \approx 1^\circ {\rm K}$.
For $\nu=4/5$, the tunneling length in the absence of any effects due 
to quantum coherence, $\ell_0$, depends on the 
impurity concentration
near the edge and the physical separation between the two
channels.  With the rough estimate, $\ell_0 \approx 10 {\rm nm}$,
and with $L = 1\mu{\rm m}$ 
and $\alpha =2$, this gives a crossover 
temperature, $T^* \approx 100 {\rm mK}$.
For $\nu = 4/3$, however, the two channels reside
in different Landau levels.  Since the Zeeman energy is significantly
less than the cyclotron energy, the two channels
will have opposite spin.
It follows that inter-channel scattering can
only occur via spin-orbit or spin flip
scattering.  The bare tunneling length $\ell_0$ should therefore
be substantially longer than for $\nu = 4/5$, and $T^*$
substantially higher.

For temperatures $T>T^*$, the neutral current $J_\sigma$ decays 
essentially to zero between
the leads, and the edge channels effectively
equilibrate, resulting in a quantized
conductance.
For $T<T^*$, however, 
full equilibration does {\it not} take place between
the contacts
and a non-universal Hall conductance given by (4.1) is expected.

While a non-universal linear conductance
is expected at low temperatures, we now argue that increasing 
the voltage can restore the quantization.   
In the following we consider the case $\nu = 4/3$.  The
results for $\nu = 4/5$ are obtained, as above, by interchanging
T and B. 
At zero temperature, $J_{T\sigma}$ decays 
according to (3.51)
due to the non-ohmic interchannel tunneling.
\begin{equation}
J_{T\sigma}(L_T) = {J_{T\sigma}(0) \over 
{\left[ 1 +   ( c_S I / I^*)^{\alpha} \right]^{1/\alpha}}}
\end{equation}
where $c_S = \alpha^{1/\alpha} G_{2S}/G_{1S}$.  Here, the
characteristic current,
\begin{equation}
I^* =   {  e k_B T_0 \over h} 
\left({\ell_0\over  L_T}\right)^{1/\alpha} = { e k_B T^*\over h}  ,
\end{equation}
sets the scale for the size of the linear response regime.
Using the above estimates we find
$I^* \approx .5 {\rm nA}$.

For $I < I^*$ there is no significant decay in the
neutral current $J_\sigma$, by the time it reaches the top
voltage contact, and
the linear conductance is not quantized. 
However,  
at higher currents, the inter-channel equilibration 
is enhanced by nonlinear tunneling.  
For $I \gg I^*$, the
neutral current at the top voltage contact is $J_{T\sigma}(L) =  \alpha^{-1/\alpha}  I^*$.  In this regime,
the measured Hall voltage may be deduced from (3.38).
\begin{equation}
\mu_T - \mu_B = {h\over \nu e}\left(  I + d_T I^* \right)  ,
\end{equation}
where $d_T = \alpha^{-1/\alpha} \nu G_{2T}/p_2 G_{1T}$.
This predicts a quantized Hall resistance
when the source-drain current $I$ is much larger than
$I^*$.
However, deviations from the quantized value, 
of order $I^*/I$, are present as a result of the incomplete
equilibration between the two edge modes.  
These deviations lead to
a slight offset of the linear I-V characteristic at high currents.
The constant $d_T$, which can be of
order $1$, depends on the structure of the voltage contact, and
can be either positive or negative.

\section{Conclusion}

As is well known in the integer quantum Hall effect, equilibration
between 
mulitple channels on
the same edge is a 
prerequisite for quantization of the Hall conductance.  
There are two sources for this equilibration:
Edge impurity scattering, and equilibration
{\it at} the electrical contacts.  
In the IQHE these can be analyzed using a free-electron
model of the edge modes.  
In this paper, we have generalized to the FQHE, 
introducing a simple kinetic equation description
of FQHE edge dynamics.
This approach allows for a unified
analysis of equilibration due to both electrical
contacts
and edge impurity scattering.
More specifically, we have introduced and analyzed
several concrete models for electrical contacts in the FQHE regime.
This allows us to describe
realistic transport geometries with multiple current and voltage contacts.

The important new feature which distinguishes FQHE edge
dynamics from the IQHE, is the presence
of edge modes which move in both directions along the edge,
such as for filling $\nu=2/3$.
In this case,
it is very difficult to equilibrate
at the electrical contacts, as we have seen in detail
by considering various specific models.
Rather, equilibration requires direct inter-channel charge
transfer, from edge impurity scattering.
This is in contrast with the IQHE, for which
multiple channels moving in the same direction
can readily be brought into equilibrium by an electrical contact.
Surprisingly, for certain quantum Hall states,
notably $\nu = 4/5$ and $\nu = 4/3$, the length scale for
equilibration between the edge channels due to impurity
scattering diverges at low temperatures.
This results in a breakdown of quantization for the Hall conductance
at low temperatures in small samples.

We hope that this work will help stimulate further experimental
exploration of mesoscopic phenomena in the fractional quantum Hall regime. 

\acknowledgements{
It is a pleasure to thank B.I. Halperin and J. Polchinski
for informative discussions.
M.P.A.F is grateful to the National Science
Foundation for support under grants No. PHY94--07194 and No. DMR--9400142.
}

\end{document}